\documentclass[aps,prb,amsmath,amssymb,twocolumn]{revtex4}
\usepackage{hyperref,amssymb,amsbsy,bm}
\pdfoutput=1

\begin{document}
\title{Spin Soret effect}
\author{Sylvain D. \surname{Brechet}}
\email{sylvain.brechet@epfl.ch}
\author{Jean-Philippe \surname{Ansermet}}
\email{jean-philippe.ansermet@epfl.ch}
\affiliation{Institute of Condensed Matter Physics, Station 3, Ecole Polytechnique F\'ed\'erale de Lausanne\,-\,EPFL, CH-1015 Lausanne,
Switzerland}

\begin{abstract}
Using a three-current model for heat, spin-up and spin-down electrons, the thermodynamics of irreversible processes predicts that a temperature gradient gives rise to a spin current under the conditions used to measure what is called the spin Seebeck effect. A diffusive current proportional to a gradient of the chemical potential is known in thermochemistry as a Soret effect or thermophoresis. 
\end{abstract}


\maketitle

\section{Introduction}

Spintronics is concerned with using the spin of the electron in novel electronic devices. The spin property adds a degree of freedom to the electron transport. Recently, it has become evident that a number of studies of spin-dependent transport included also heat transport and the term ``spin caloritronic'' was coined~\cite{Bauer:2010}. While the charge, spin and heat currents were considered by spintronics founders~\cite{Johnson:1985}, much attraction to the inclusion of heat in spin-dependent transport stems from the discovery of the so-called \textit{spin Seebeck effect}~\cite{Uchida:2008,Jarowski:2010}.
 
The Seebeck effect is known in the theory of non-equilibrium thermodynamics as a cross-effect involving two generalized forces, a temperature gradient and an electrostatic potential gradient~\cite{Kempers:2001}. Here we review another cross-effect, the Soret effect. It is closely related to the Seebeck effect and we show how it may apply to spin caloritronics.

Charles Soret observed that a tube containing a mixture of two salts, where one end of the tube was maintained cold and the other hot, presented different salt concentrations at both ends~\cite{Soret:1879}. Since then, the diffusive current of one substance with respect to another, driven by a temperature gradient, has been known as the \textit{Soret effect} or \textit{thermophoresis}. In liquids, it is usually demonstrated using two plates, where the top one is maintained hot and the bottom one cold in order to avoid convection. After a short while, the balance between the heat current and the diffusion flux leads the system to a stationary state~\cite{Fitts:1962}. Thermodiffusion is more commonly observed with aerosols in which the effect is quite intuitive : the air molecules have on average a higher velocity on the higher temperature side of any particle in suspension, thus pushing it to the lower temperature side.

\section{Three current model}

In order to examine the Soret effect in the context of spin transport, we consider that the charge carriers are of two types, up spins and down spins,   labeled with a subscript $(+)$ and $(-)$, respectively. Then, we use a three current model involving a heat current density, and the two charge current densities. We use the notation of a previous work~\cite{Gravier:2006}. A similar three-current model was recently used by~\cite{Slachter:2010}. The three diffusive currents are the entropy current density $\mathbf{j}_s$ and the electric current densities $\mathbf{j}_+$ and $\mathbf{j}_-$ of the spin up and spin down carriers. Thermodynamics of irreversible processes implies that there are linear relations between these current densities and their respective generalized forces, which are the Onsager reciprocity relations~\cite{Onsager:1931}. These relations relate the heat current $\mathbf{j}_s$ and the charge current densities $\mathbf{j}_+$ and $\mathbf{j}_-$ respectively to the gradient of temperature $\bm{\nabla}\,T$, and the gradients of the respective electrochemical potential of the charge carriers, $\bm{\nabla}\mu_+$ and $\bm{\nabla}\mu_-$ according to, 
\begin{equation}\label{Onsager 1}
\begin{pmatrix} 
\mathbf{j}_s\\ 
\mathbf{j}_+\\
\mathbf{j}_-
\end{pmatrix}
= -
\begin{pmatrix} 
\mathcal{L}_{ss} & \mathcal{L}_{s+} & \mathcal{L}_{s-}\\ 
\mathcal{L}_{+s} & \mathcal{L}_{++} & \mathcal{L}_{+-}\\
\mathcal{L}_{-s} & \mathcal{L}_{-+} & \mathcal{L}_{--}
\end{pmatrix}
\begin{pmatrix} 
\bm{\nabla}\,T\\ 
\bm{\nabla}\mu_+\\
\bm{\nabla}\mu_-
\end{pmatrix}
\ .
\end{equation}
For convenience, we adopt the notation of Valet and Fert~\cite{Valet:1993} and express the electrochemical potentials $\mu_{\pm}$ as,
\begin{equation}\label{mu el-ch}
\mu_{\pm}=\mu_0\pm\Delta\mu + qV\ ,	
\end{equation}
where $q$ is the charge per carrier, $V$ the applied electrostatic potential and $\mu_0$ the mean chemical potential. In our discussion, we do not consider purely chemical effects, so $\bm{\nabla}\mu_0=0$. We identify now the physical meaning of the Onsager coefficients based on phenomenological considerations. 

The Onsager coefficients $\mathcal{L}_{+-}$ and $\mathcal{L}_{-+}$ describe spin mixing~\cite{Ansermet:2008}. These coefficients are important when describing the magnetic field dependence of the thermoelectric power~\cite{Tsyplyatyev:2006}. Here we neglect them as they simply modify slightly the coefficients of our final result, i.e. 
\begin{equation}\label{neglect I}
\mathcal{L}_{+-}=0\qquad\text{and}\qquad\mathcal{L}_{-+}=0\ .	
\end{equation}

The spin-dependent electric conductivities $\sigma_{\pm}$ of each spin current are determined in the absence of a temperature gradient, i.e. $\bm{\nabla}\,T=0$ and chemical potential gradient, i.e. $\bm{\nabla}\mu_{\pm}=0$, which then implies that~\eqref{Onsager 1} reduces to,
\begin{equation}\label{cond + -}
\mathbf{j}_{\pm}=-\,\mathcal{L}_{\pm\pm}\,q\bm{\nabla}\,V=\sigma_{\pm}\mathbf{E}\ ,	
\end{equation}
where $\mathbf{E}=-\bm{\nabla}\,V$ is the applied electric field. Thus, the Onsager coefficients $\mathcal{L}_{\pm\pm}$ are recast in terms of the spin-dependent electric conductivities respectively as,
\begin{equation}\label{cond Onsager}
\mathcal{L}_{++}=\frac{\sigma_{+}}{q}\qquad\text{and}\qquad\mathcal{L}_{--}=\frac{\sigma_{-}}{q}\ .	
\end{equation}

The spin-dependent Seebeck coefficients, $\varepsilon_{\pm}$, defined as,
\begin{equation}\label{spin Seebeck I}
\varepsilon_{\pm}=-\,\frac{1}{q}\frac{\bm{\nabla}\mu_{\pm}}{\bm{\nabla}\,T}\ ,	
\end{equation}
are compared to virtual experiments in which one imposes one spin current density to vanish, i.e. $\mathbf{j}_{\pm}=0$, which then implies that the corresponding relations in~\eqref{Onsager 1} reduce to,
\begin{equation}\label{Seebeck virtual exp}
\mathcal{L}_{\pm s}\,\bm{\nabla}\,T+\frac{\sigma_{\pm}}{q}\,\bm{\nabla}\mu_{\pm} = 0\ .	
\end{equation}
Thus, the Onsager coefficients $\mathcal{L}_{\pm s}$ are recast in terms of the spin-dependent electric conductivities $\sigma_{\pm}$ and the spin-dependent Seebeck coefficients $\varepsilon_{\pm}$ respectively as,
\begin{equation}\label{cond Onsager II}
\mathcal{L}_{+s}=\sigma_{+}\varepsilon_{+}\qquad\text{and}\qquad\mathcal{L}_{-s}=\sigma_{-}\varepsilon_{-}\ .	
\end{equation}

The Onsager reciprocity relations state that the coefficients $\mathcal{L}_{ij}$ and $\mathcal{L}_{ji}$, where $i,j=\{s,+,-\}$, describe reverse processes under a time reversal. Since the spin change sign under a time inversion, up spins correspond to down spins in reverse processes and vice versa. Thus the Onsager coefficients $\mathcal{L}_{s\pm}$ are related to the reverse Onsager coefficients $\mathcal{L}_{\mp s}$ by,   
\begin{equation}\label{cond Onsager III}
\mathcal{L}_{s+}=\mathcal{L}_{-s}\qquad\text{and}\qquad\mathcal{L}_{s-}=\mathcal{L}_{+s}\ .	
\end{equation}

The thermal conductivity $\kappa$ is determined in the absence of spin-depend current, i.e. $\mathbf{j}_+=\mathbf{j}_-=0$. Using the relations~\eqref{spin Seebeck I},~\eqref{cond Onsager II} and~\eqref{cond Onsager III}, the thermal equation in~\eqref{Onsager 1} yields,
\begin{equation}\label{cond eq 2}
\mathbf{j}_{s}=-\left[\mathcal{L}_{ss}-\left(\sigma_+ + \sigma_-\right)\varepsilon_+\varepsilon_-\right]\bm{\nabla}\,T=-\frac{\kappa}{T}\bm{\nabla}\,T\ .	
\end{equation}
Thus, the Onsager coefficient $\mathcal{L}_{ss}$ is recast in terms of the thermal conductivity coefficient $\kappa$, the spin-dependent electric conductivities $\sigma_{\pm}$ and the spin-dependent Seebeck coefficients $\varepsilon_{\pm}$ respectively as,
\begin{equation}\label{thermal conductivity}
\mathcal{L}_{ss} = \frac{\kappa}{T} + \left(\sigma_+ + \sigma_-\right)\varepsilon_+\varepsilon_- = \kappa'\ ,	
\end{equation}
where $\kappa'$ is the ratio of the thermal conductivity divided by the temperature. Using the analytic expressions for the Onsager coefficents~\eqref{neglect I},~\eqref{cond Onsager},~\eqref{cond Onsager II},~\eqref{cond Onsager III} and~\eqref{thermal conductivity}, the Onsager relations~\eqref{Onsager 1} become,
\begin{equation}\label{Onsager 2}
\begin{pmatrix} 
\mathbf{j}_s\\ 
\mathbf{j}_+\\
\mathbf{j}_-
\end{pmatrix}
= -
\begin{pmatrix} 
\kappa' & \sigma_-\varepsilon_- & \sigma_+\varepsilon_+\\ 
\sigma_+\varepsilon_+ & \frac{\sigma_+}{q} & 0\\
\sigma_-\varepsilon_- & 0 & \frac{\sigma_-}{q}
\end{pmatrix}
\begin{pmatrix} 
\bm{\nabla}\,T\\ 
\bm{\nabla}\mu_+\\
\bm{\nabla}\mu_-
\end{pmatrix}
\ .
\end{equation}

At this point, it is physically useful to define the electric current density $\mathbf{j}$ and the spin-dependent electric polarisation current density $\mathbf{j}_p$ as the sum and the difference of the spin-dependent current densities $\mathbf{j}_+$ and $\mathbf{j}_-$ respectively, i.e.
\begin{equation}
\mathbf{j}=\mathbf{j}_+ + \mathbf{j}_-\qquad\text{and}\qquad\mathbf{j}_p=\mathbf{j}_+ - \mathbf{j}_-\ .
\end{equation}
Similarly, we define the effective electric conductivity $\sigma$ and the spin-dependent polarisation conductivity $\sigma_p$ as,
\begin{equation}
\sigma=\sigma_+ + \sigma_-\qquad\text{and}\qquad\sigma_p=\sigma_+ - \sigma_-\ .
\end{equation}

Spin relaxation occurs near the electrodes. At distances much greater than the diffusion length the spin channels reach equilibrium, which implies that $\bm{\nabla}\left(\Delta\mu\right)=0$ and from~\eqref{mu el-ch}, it follows that $\bm{\nabla}\mu\equiv\bm{\nabla}\mu_{+}=\bm{\nabla}\mu_{-}$.

In this bulk limit, we define the effective spin Seebeck coefficient $\varepsilon$ and the polarisation spin Seebeck coefficient $\varepsilon_p$ respectively as,
\begin{equation}\label{coef Seebeck eff}
\varepsilon=-\,\frac{1}{q}\frac{\bm{\nabla}\mu}{\bm{\nabla}\,T}\quad\!\!\text{for}\quad\!\!\mathbf{j}=0\,\,\ ,\quad\varepsilon_p=-\,\frac{1}{q}\frac{\bm{\nabla}\mu}{\bm{\nabla}\,T}\quad\!\!\text{for}\quad\!\!\mathbf{j}_p=0\ .
\end{equation}

The Onsager relations~\eqref{Onsager 2} imply for $\mathbf{j}=0$ and $\mathbf{j}_p=0$ respectively that, 
\begin{align}\label{coef Seebeck eff 2}
\begin{split}
&\left(\sigma_{+}\varepsilon_{+} + \sigma_{-}\varepsilon_{-}\right)\bm{\nabla}\,T+\frac{\sigma_{+} + \sigma_{-}}{q}\,\bm{\nabla}\mu = 0\ ,\\
&\left(\sigma_{+}\varepsilon_{+} - \sigma_{-}\varepsilon_{-}\right)\bm{\nabla}\,T+\frac{\sigma_{+} - \sigma_{-}}{q}\,\bm{\nabla}\mu = 0\ .
\end{split}
\end{align}

By comparing the relations~\eqref{coef Seebeck eff} and\eqref{coef Seebeck eff 2}, the analytical expressions for the spin Seebeck coefficients $\varepsilon$ and $\varepsilon_p$ are respectively found to be,
\begin{equation}\label{coef Seebeck eff 3}
\varepsilon=\frac{\sigma_{+}\varepsilon_{+} + \sigma_{-}\varepsilon_{-}}{\sigma_{+} + \sigma_{-}}\qquad\text{and}\qquad\varepsilon_p=\frac{\sigma_{+}\varepsilon_{+} - \sigma_{-}\varepsilon_{-}}{\sigma_{+} - \sigma_{-}}\ .
\end{equation}

Finally, the Onsager matrix~\eqref{Onsager 2} can be recast as,
\begin{equation}\label{Onsager 3}
\begin{pmatrix} 
\mathbf{j}_s\\ 
\mathbf{j}\\
\mathbf{j}_p
\end{pmatrix}
= -
\begin{pmatrix} 
\kappa' & q\,\sigma\varepsilon & q\,\sigma_p\varepsilon_p\\ 
\sigma\varepsilon & \sigma & \sigma_p\\
\sigma_p\varepsilon_p & \sigma_p & \sigma
\end{pmatrix}
\begin{pmatrix} 
\bm{\nabla}\,T\\ 
\bm{\nabla}\,V\\
\frac{1}{q}\bm{\nabla}\left(\Delta\mu\right)
\end{pmatrix}
\ .
\end{equation}

In chemistry, the Soret effect refers to the contribution to the diffusive current of the solute with respect of solvent~\cite{Fitts:1962}, which is due to the temperature gradient. By analogy, the spin Soret effect refers to the contribution to polarisation current $\mathbf{j}^s_p$ due to the temperature gradient $\bm{\nabla}\,T$. 

By defining the spin Soret coefficient $\Sigma$ as,
\begin{equation}\label{Soret coef}
\mathbf{j}^s_p=-\,\Sigma\,\bm{\nabla}\,T\ ,	
\end{equation}
it follows from~\eqref{Onsager 3}, that the spin Soret coefficient $\Sigma$ is given by,
\begin{equation}\label{Soret coef 2}
\Sigma=\sigma_p\varepsilon_p\ .	
\end{equation}

\section{Spin Seebeck effect}

We analyse now the spin Seebeck experiment. The platinum electrodes are assumed to produce a very strong spin relaxation, so that the experiment can be modeled with $\Delta\mu=0$ at both electrodes. Then, the diffusion equation for $\Delta\mu$ implies $\bm{\nabla}\left(\Delta\mu\right)=0$. As in any Seebeck measurement, there is no electric current, i.e. $\mathbf{j}=0$. Under these conditions, using~\eqref{coef Seebeck eff}, the spin polarisation current density $\mathbf{j}_p$ in~\eqref{Onsager 3} is found to be proportional to the temperature gradient and reduces to the elegant and simple expression,
\begin{equation}\label{curr pol 1}
\mathbf{j}_p=-\,\sigma_p\left(\varepsilon_p - \varepsilon\right)\bm{\nabla}\,T\ .	
\end{equation}

The spin-dependent electric conductivities $\sigma_{\pm}$ and the spin-dependent spin Seebeck coefficients $\varepsilon_{\pm}$ can be written as,
\begin{equation}\label{coef Seebeck expansion}
\sigma_{\pm}=\frac{\sigma}{2}\left(1\pm\beta\right)\ ,\qquad\text{and}\qquad\varepsilon_{\pm}=\varepsilon\left(1\pm\eta\right)\ .
\end{equation}
Thus, to first order in $\beta$ and $\eta$,
\begin{equation}\label{rel first od}
\sigma_p=\beta\sigma\ ,\qquad\text{and}\qquad\varepsilon_p=\varepsilon\left(1 + \frac{\eta}{\beta}\right)\ ,	
\end{equation}
and the expression~\eqref{curr pol 1} for the spin polarisation current density $\mathbf{j}_p$ reduces to,
\begin{equation}\label{curr pol 2}
\mathbf{j}_p=-\,\sigma\varepsilon\left(\eta - \beta\right)\bm{\nabla}\,T\ .	
\end{equation}

In~\cite{Uchida:2008}, a spin polarisation current density $\mathbf{j}_p$ is detected by the inverse spin Hall effect occurring in the platinum electrodes. Uchida et al.~\cite{Uchida:2009} deduced a spin current from a two-current model by assuming that the temperature derivative of the chemical potential was spin-dependent, which is quite different from the present description. 

There are many systems where two types of carriers (of charge as well as of entropy) are invoked. For example, in reference to the observation of a spin Seebeck effect in insulating materials~\cite{Uchida:2010}, it has been conjectured that the difference between the effective temperature of magnons and that of phonons drives spin pumping into the electrons~\cite{Xiao:2010}. From a mere thermodynamic standpoint, a drift of the magnons~\cite{Kajiwara:2010} toward the cold side of the sample is a form of thermophoresis. Phonon drag can also be thought of as a cross-effect between phonons and electrons~\cite{Miele:1998}. It has been suggested that it may also play a role in the spin Seebeck effect~\cite{Adachi:2010}.

As always with the thermodynamics of irreversible processes, the Onsager matrix expresses the existence of relationships among the gradients of the extensive state variables and their associated currents, however this phenomenological approach cannot address the micro-mechanisms responsible for the values of the matrix elements. 

\section{Conclusion}

In summary, we consider the spin Seebeck effect as an experiment characterised by zero effective electric current and the equilibrium of the chemical potential at the electrodes. We show that a three-current model of heat, spin-up and spin-down currents predicts the spin polarisation current to be proportional to the temperature gradient, corresponding to a spin Soret effect. 



\bibliography{references} 

\end{document}